\documentclass[reprint,superscriptaddress,amsmath,amssymb,longbibliography,floatfix]{revtex4-1}

\usepackage{graphicx}
\usepackage{dcolumn}
\usepackage{bm}
\usepackage{hyperref}

\newcommand{\ilm}{Univ Lyon, Univ Claude Bernard Lyon 1, CNRS, Institut Lumi\`ere Mati\`ere, F-69622, VILLEURBANNE, France}
\newcommand{\uzh}{Department of Chemistry, Universit\"at Z\"urich, 8057 Z\"urich, Switzerland}
\newcommand{\iuf}{Institut Universitaire de France (IUF), 1 rue Descartes, 75005 Paris, France}

\begin{document}


\title{Fast Increase of Nanofluidic Slip in Supercooled Water: the Key Role of Dynamics}

\author{Cecilia Herrero}
\affiliation{\ilm}
\author{Gabriele Tocci}
\affiliation{\uzh}
\author{Samy Merabia}
\affiliation{\ilm}
\author{Laurent Joly}
\email{laurent.joly@univ-lyon1.fr}
\affiliation{\ilm}
\affiliation{\iuf}

\date{\today}

\begin{abstract}
Nanofluidics is an emerging field offering innovative solutions for energy harvesting and desalination. The efficiency of these applications depends strongly on liquid-solid slip, arising from a favorable ratio between viscosity and interfacial friction. Using molecular dynamics simulations, we show  that wall slip increases strongly when water is cooled below its melting point. For water on graphene, the slip length is multiplied by up to a factor of five and reaches $230$\,nm at the lowest simulated temperature, $T \sim 225$\,K; experiments in nanopores can reach much lower temperatures and could reveal even more drastic changes. The predicted fast increase in water slip can also be detected at supercoolings reached experimentally in bulk water, as well as in droplets flowing on anti-icing surfaces. We explain the anomalous slip behavior in the supercooled regime by a decoupling between viscosity and bulk density relaxation dynamics, and we rationalize the wall-type dependency of the enhancement in terms of interfacial density relaxation dynamics. By providing fundamental insights on the molecular mechanisms of hydrodynamic transport in both interfacial and bulk water in the supercooled regime, this study is relevant to the design of anti-icing surfaces and it also paves the way to explore new behaviors in supercooled nanofluidic systems.
\end{abstract}

\maketitle

\paragraph*{Introduction--}

Nanofluidics, i.e. the study of fluidic transport at nanometer scales, has emerged as a new and interesting field in the past few decades due to novel behaviors associated to this length scale \cite{Koga2001,schoch2008,bocquet2010} -- e.g. dielectric anomalies of confined water \cite{Fumagalli2018} or intriguing ionic transport \cite{Feng2016a,Kavokine2019,Mouterde2019}, 
with promising applications related to new 2D materials  
such as the development of sustainable energies \cite{Siria2017,Marbach2019,Xu2020}. 
As confinement increases, interfacial properties have an increasingly important role. 
An interfacial characteristic of special concern at the nanoscale is 
the existence of a velocity jump $\Delta v$ (`slippage') at the liquid-solid interface \cite{lauga2005,neto2005,bocquet2007}. The simplest approach to describe slip, initially proposed by Navier \cite{Navier1823}, is to consider that the viscous shear stress $\tau$ in the liquid at the wall is proportional to the velocity jump, $\tau = \lambda \Delta v$, where 
$\lambda$ is the liquid-solid friction coefficient. 

Because reducing friction is key to improving the performance of nanofluidic systems, an intensive experimental effort has been undertaken during the recent years to characterize the ultra-low liquid-solid friction of new 2D materials and their derivative \cite{Secchi2016,Yang2017,Tunuguntla2017,Xie2018}. 
On the modeling side, 
several efforts have been pursued in order to understand the molecular mechanisms that control friction, with special interest on the discussion of the relation between the friction coefficient and the time autocorrelation of the force exerted by the liquid on the wall \cite{bocquet1994,petravic2007,hansen2011,huang2014,oga2019green,espanol2019,Nakano2020,Straube2020}. Further work has been performed to study the impact on friction of different wall features such as wettability  \cite{BB1999faraday,sendner2009}, roughness \cite{gu2011}, crystallographic orientation \cite{wagemann2017}, electronic structure  \cite{Tocci2014,Tocci2020,Xie2020}, or electrostatic interactions \cite{govind2019}. 
Yet a large number of questions with regard to the interface properties, such as its viscoelastic or purely viscous nature \cite{cross2018,omori2019,Grzelka2020} or the possible link with its interfacial thermal transport equivalents via wall's wetting properties \cite{barrat2003kapitza,caplan2014,giri2014}, remain open nowadays, limiting the perspectives for a rational search of optimal interfaces.

Among all fluids, the study of water has always been of special concern for scientists from a broad variety of research fields  \cite{robinson1996,franks2007,ball2008}.  
Its interest not only lies on its ubiquitous nature but also on its many thermodynamic and dynamic anomalies 
like, among others, the non-monotonous temperature dependence of its isothermal compressibility and density 
\cite{debenedetti2003,gallo2016}. These anomalies are enhanced when water is driven to its supercooled regime (i.e. the range of temperatures below the freezing point where water keeps its liquid state), 
making this regime ideal to test and refine our current understanding of water. 
In particular, the temperature dependence of the bulk transport properties of supercooled water has been explored both numerically and experimentally over the last decade \cite{dehaoui2015,montero2018}, considering especially the connection between viscosity and structural relaxation times \cite{yamamoto1998,chen2006,xu2009,shi2013,Kawasaki2017}. 

The temperature evolution of supercooled water under confinement
has also been the subject of intensive experimental research \cite{chen2006,buchsteiner2006water,xu2009,cerveny2010dynamics,Cerveny2016,Kaneko2018}. 
Broadband dielectric spectroscopy, nuclear magnetic
resonance, as well as neutron scattering experiments
have successfully probed water confined
in pores with sub-nm radii at temperatures as low as about 130\,K,
in order to connect the
dynamical behavior of supercooled confined water
to that of bulk water in the so-called no-man's land (150\,K to 230\,K) \cite{Cerveny2016}.
At temperatures above the no-man's land, 
marked differences have been found in the
time relaxation of supercooled water under confinement
compared to bulk water, suggesting that the
interfacial water dynamics, and thus water friction,
may play an important role.
However, the temperature evolution of water friction in the liquid and supercooled regime remains unclear nowadays. 
Besides achieving a better understanding of  interfacial and nanoconfined water
dynamics and phase behavior under supercooling, such a knowledge would be instrumental e.g. for the development of innovative nanofluidic systems working in the supercooled regime, and would provide fundamental insight on recent experimental work on anti-icing surfaces \cite{Mishchenko2010,Jung2012,Kreder2016}.

In that context, we report a study in which the temperature dependence of liquid-solid friction and of bulk liquid viscosity are examined in detail, in connection to the dynamical behavior of interfacial liquids in the supercooled regime. To this end we perform extensive molecular dynamics (MD) simulations of water and methanol on graphene and generic Lennard-Jones surfaces.
We find that whilst the liquid-solid friction coefficient and the viscosity follow the same fundamental laws and are almost proportional to each other in the liquid state, 
their behavior strikingly differ in the supercooled regime. 
As a result, the slip length -- defined as the ratio between the viscosity and the friction coefficient -- increases fast for water as soon as it goes below its melting point, at temperatures that were
recently accessible  
in experiments on bulk water \cite{dehaoui2015}, in
water droplets \cite{Jung2012} and under confinement \cite{chen2006}; on graphene, we report a twofold enhancement at $\sim 240$\,K, and up to a fivefold enhancement at  225\,K, reaching $\sim 230$\,nm.
Our analysis reveals that the dynamics of interfacial water, specifically the time relaxation of the interfacial density fluctuations, is the most important factor governing the temperature behavior of liquid-solid friction and slip. This fundamental mechanistic insight sheds new light on the general molecular mechanisms underlying water slip.

\paragraph*{MD Simulations--} 

\begin{figure}
    \centering
    \includegraphics[width=0.8\linewidth]{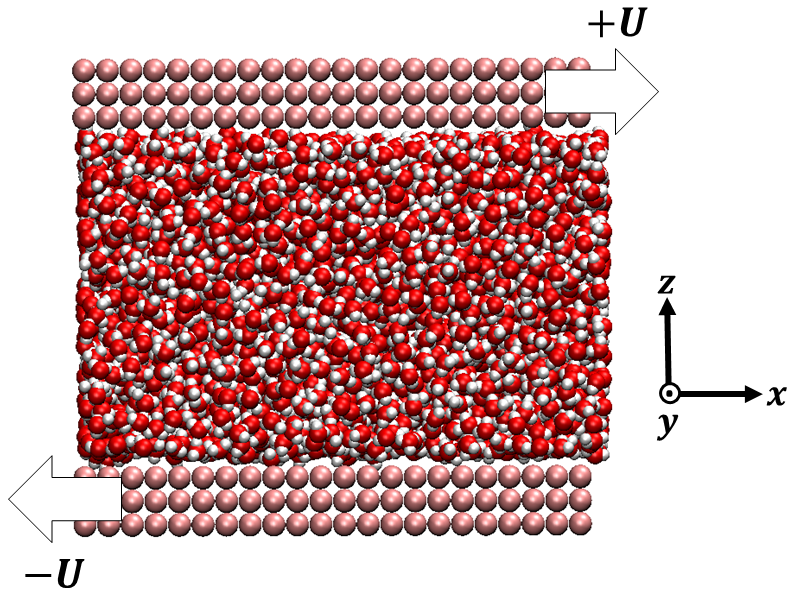}
    \caption{Modelled system constituted by a confined fluid between two planar solid walls. The snapshot corresponds to TIP4P/2005 water enclosed by LJ walls. The arrows indicate the shear velocity $U$ directions by which the system is driven out of equilibrium for the shear flow measurements.}
    \label{fig:system}
\end{figure}

All the simulations were carried out with the LAMMPS package \cite{plimpton1995}. The confined system consisted in a fluid -- TIP4P/2005 water \cite{abascal2005} or methanol (MeOH) \cite{schnabel2007,guevara2011} -- between two parallel walls -- graphene, or a generic hydrophobic wall made of Lennard-Jones (LJ) particles -- with periodic boundary conditions applied in the directions parallel to the walls  (Fig.~\ref{fig:system}), see details in the Electronic Supplementary Information (ESI). The surfaces were characterized by contact angles of $\theta \sim 134^{\circ}$ for water-LJ walls, $\theta \sim 80^{\circ}$ for water-graphene, $\theta \sim 100^{\circ}$ for MeOH-LJ walls and $\theta \sim 0^{\circ}$ for MeOH-graphene.

The wall dimensions were $L_x = L_y = 58.92$\,\AA{} for the LJ wall, and $L_x = 56.57$\,\AA{}, $L_y = 58.92$\,\AA{} for graphene. The pressure was set to 1\,atm by using the top wall as a piston during a preliminary run; the vertical height was then obtained by fixing the top wall at its equilibrium position for the given pressure and it corresponded to $H \sim 40$\,\AA{} for water and $H \sim 90$\,\AA{} for MeOH. The temperature $T$ was varied between $225$ and $360$\,K, by applying a Nos\' e-Hoover thermostat to the liquid (only along the directions perpendicular to the flow for non-equilibrium simulations). Equivalent results were obtained for different damping times, and with a Berendsen thermostat.

To measure the hydrodynamic transport coefficients we performed non-equilibrium molecular dynamics (NEMD) simulations, applying a constant shear velocity $U$ to the walls in opposite $x$ directions for each wall (see Fig.~\ref{fig:system}), producing a linear velocity profile far from the wall. The friction coefficient was measured from the ratio between the shear stress $\tau$ and the velocity jump at the interface $\Delta v$ -- defined at the effective wall position $z_s$ \cite{herrero2019}: $\lambda = \tau/ \Delta v$, and the viscosity was measured from the ratio between the shear stress and the bulk shear rate, $\eta = \tau / (\partial_z v_x)$, see the ESI for details. 

Both interfacial and bulk equations can be combined in the so-called partial slip boundary condition \cite{bocquet2007}, 
\begin{equation}
    \Delta v = \frac{\eta}{\lambda} \,\partial_z v_x \Big|_{z=z_s} = b \,\partial_z v_x \Big|_{z=z_s},
    \label{eq:Navier-BC}
\end{equation}
defining the slip length $b := \eta/\lambda$. 
Viscosity and friction have been measured for $3$ different shear velocities for each temperature, $U \in [1, 70]$\,m/s in order to verify that our measurements where performed in the linear response regime. 
For a given shear velocity, $3$ independent simulations were run and we measured the shear stress at the top and bottom walls for each of them. 
Overall, $18$ independent measurements were taken for a given $T$ and the error bars in this article correspond to the statistical error within $95\%$ of confidence level.

\paragraph*{Results and discussion--}

We first computed the shear viscosity $\eta$ from NEMD with LJ walls to test the applicability of the different temperature dependence laws, Vogel-Tammann-Fulcher (VTF), Speedy-Angell (SA) and B\"{a}ssler (B) laws, see the ESI. For TIP4P/2005 we find good agreement between our data and the experimental ones \cite{dehaoui2015, hallett1963}, as well as previous MD simulations with the TIP4P/2005 and TIP4P/2005f water models \cite{guevara2011,Markesteijn2012,guillaud2017decoupling}. Our viscosity measurements are best described by VTF law 
(see the ESI). For MeOH simulations viscosity's temperature dependence is weaker than for water. The results are in good agreement with previous work \cite{guevara2011} and they are well described by an Arrhenius law.

\begin{figure}
    \centering
    \includegraphics[width=0.8\linewidth]{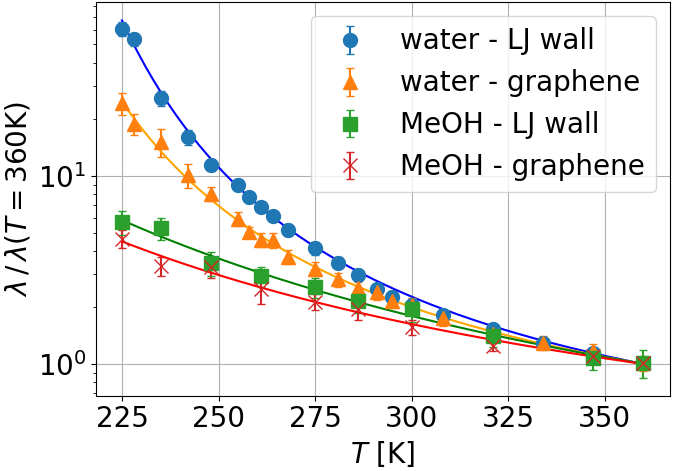}
    \caption{Friction's temperature dependence results normalized by the value at $360$\,K for each fluid and wall, in order to highlight the similar temperature evolution for a given liquid regardless of the wall type. Blue dots correspond to water with LJ walls, orange triangles to water with graphene walls, green squares to MeOH with LJ walls and red crosses to MeOH with graphene walls. Continuous lines are the respective VTF (for water) and Arrhenius (for MeOH) fits.} 
    \label{fig:Friction_Tratio}
\end{figure}

We then proceeded to explore temperature effects on friction. 
For each fluid, when varying the wall type, we already saw a difference at a given temperature in the absolute value of $\lambda$, being more than one order of magnitude smaller for graphene than for LJ walls (see the ESI). This effect has already been measured and discussed in previous work \cite{falk2010,kannam2013,Tocci2014,Sam2020} and it is due to the extreme smoothness of graphene. 
Additionally, in Fig.~\ref{fig:Friction_Tratio} one can see that the temperature dependence changes with the fluid, but for a given fluid, depends weakly on the  wall type. 
Interestingly, the temperature dependence of $\eta$ and $\lambda$ can be fitted by the same laws (VTF for TIP4P/2005 and Arrhenius for MeOH, corresponding to continuous lines in Fig.~\ref{fig:Friction_Tratio}), although with different parameters. 

\begin{figure}
    \centering
    \includegraphics[width=\linewidth]{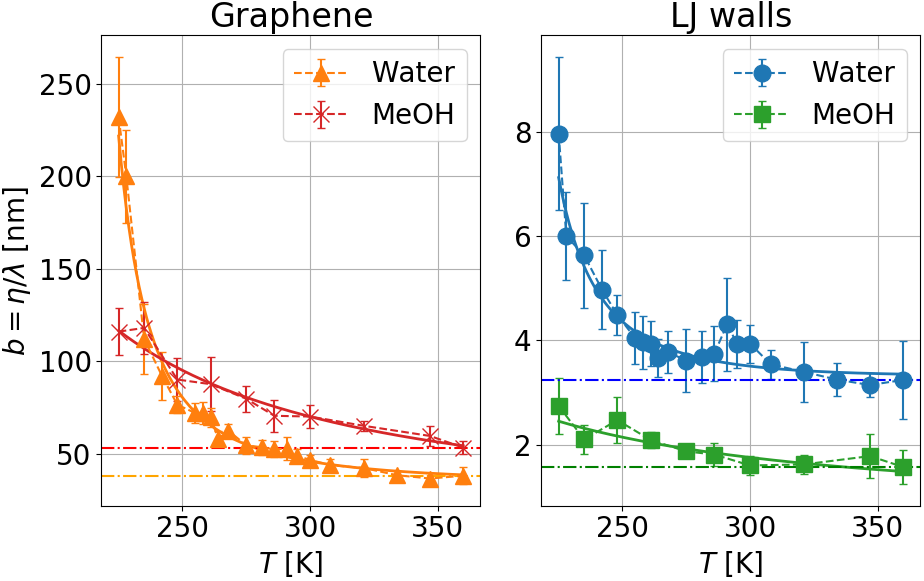}
    \caption{Temperature dependence of the slip length, $b=\eta/\lambda$, with the same symbols as Fig.~\ref{fig:Friction_Tratio}. Dash-dotted lines are guide-to-the-eye for a constant $b$ value. One can see a small temperature variation for the highest temperatures (indicating that $\eta$ and $\lambda$ evolve in similar ways), while the slip length increases significantly when decreasing the temperature for the lowest $T$s, in the supercooled regime.} 
    \label{fig:slip_length}
\end{figure}

We can go further in exploring the relation between $\eta$ and $\lambda$ by plotting the slip length $b$ given by the ratio between both transport coefficients, see Eq.~\eqref{eq:Navier-BC}. In Fig.~\ref{fig:slip_length} one can see that for a wide range of high temperatures, where the systems are in the stable liquid state, 
$\eta$ and $\lambda$ vary together with $T$, so their ratio (or equivalently the slip length) is roughly constant. 
Specifically, for MeOH (which remains above its melting point for the whole range of simulated temperatures), $b$ increases slowly and regularly when $T$ decreases; this indicates a slightly weaker temperature dependence of friction as compared to viscosity.  
In contrast, for water, $b$ starts to increase very fast when the temperature decreases below the melting point, indicating a much weaker temperature dependence of friction as compared to viscosity, only in the supercooled regime. 
The biggest temperature effect on $b$ is observed for water and graphene walls, where it grows by a factor of $5$ from the highest to the lowest simulated temperature (225\,K), reaching a maximum value of $\sim 230$\,nm.
Although experiments of interfacial slip in supercooled water have not yet appeared, we envision that experimental verification of our results may be within reach of capillary flow measurements of water confined between graphene/silica nanochannels \cite{Xie2018}, considering that nuclear magnetic resonance and
neutron scattering experiments of water confined in graphite oxide  and silica nanopores have explored water dynamics down to 130\,K and 220\,K, respectively \cite{buchsteiner2006water,cerveny2010dynamics,chen2006}. 
Additionally, recent microscopy studies have investigated the dynamical behavior of supercooled water  down to 230\,K using  polyesterene spheres suspended in  water \cite{dehaoui2015} and have studied the anti-icing behavior of water droplets sliding on of nanopatterned surfaces around 258\,K \cite{Jung2012}.
As we have seen in our simulations, at temperatures between 225 and 270\,K, water slippage is largely affected by the type of interface, a result consistent with
the experimental observation that the water time relaxation in nanopores
is strongly dependent on the type of confinement. Instead, experiments under  in  no-man's land report a universal dynamical behaviour in confined water \cite{Cerveny2016}. Thus, future measurements of water slippage in different nanopores at lower temperatures than those explored here might elucidate whether or not slippage depends on the type of interface below 225\,K.
Enhanced water slippage under supercooling may also have a direct implication for
the development of icephobic surfaces,
as it may favour droplet 
condensation and ice removal from surfaces.

Two main questions remain then to be understood. First, what is the main physical parameter that controls the temperature evolution of the friction coefficient observed in Fig.~\ref{fig:Friction_Tratio}? Second, why bulk and interface have a similar temperature dependence at high temperatures and why they don't at the lower ones (Fig.~\ref{fig:slip_length})? Because the effect of temperature is larger for water, from now on we will focus on this liquid to address these questions.

In order to better understand the molecular mechanisms that control friction temperature dependence, $\lambda$ can be related 
to the autocorrelation of the equilibrium force at the interface through a Green-Kubo formula \cite{bocquet1994,bocquet2013}: 
\begin{equation}
    \lambda = \frac{1}{A k_B T} \int_0^{\infty} \langle F(t)F(0) \rangle \mathrm{d}t,
    \label{eq:fric_GK}
\end{equation}
where $A$ is the surface area, $k_B$ the Boltzmann's constant, $T$ the temperature and $F$ the force applied by the fluid on the wall. This expression can be decomposed as a product of static (``STAT'') and dynamical (``DYN'') terms of the form -- see Ref.~\citenum{BB1999faraday} and the ESI:
\begin{equation}
\begin{split}
     \lambda 
     &\equiv \lambda_{\mathrm{STAT}} \cdot \lambda_{\mathrm{DYN}}, 
     \text{ with }\\
     \lambda_{\mathrm{STAT}} &\approx S(q_{\parallel})  \int_0^{\infty} \mathrm{d}z \, \rho(z) f_{q_{\parallel}}^2(z)  \\
     \lambda_{\mathrm{DYN}} &\approx  \frac{\tau_\rho}{2 k_B T} 
\end{split}
     \label{eq:fric_dyn}
\end{equation}
where $S(q_{\parallel})$ is the 2D structure factor in the contact layer, evaluated at the shortest wave vector of the solid surface $q_{\parallel}$, 
$\rho(z)$ is the fluid number density, 
$f_{q_{\parallel}}(z)$ is the force corrugation and 
$\tau_{\rho}$ is the density relaxation time defined as the integral of the intermediate scattering function in the contact layer taken at $q_{\parallel}$: $\tau_{\rho} = \int_{0}^{\infty} \mathrm{d}t F(q_{\parallel},t)$. The contact layer was defined as the liquid region between the wall and the first minimum of the liquid's density profile. 
Note that we included the $1/(k_\text{B}T)$ term of the Green-Kubo integral in the dynamical part; we will come back to that choice later. 
Regarding the static terms in Eq.~\eqref{eq:fric_dyn}, 
we found that $S(q_{\parallel})$ remained constant with temperature for both graphene and LJ walls (see the ESI). 
The main static contribution to friction $T$ dependence comes from the integral in Eq.~\eqref{eq:fric_dyn}. We used for $f_{q_{\parallel}}(z)$ the analytical expression derived in Ref.~\citenum{steele1973} for LJ walls and the measurements in Ref.~\citenum{falk2010} for graphene (as detailed in the ESI). 
For both surfaces, the integral remains constant at low temperatures, and then increases by at most a factor of 2 at higher temperatures. Therefore, 
this temperature behavior does not explain the exponential decrease for increasing temperature observed for friction. 
It is only left to check the dynamical contribution from Eq.~\eqref{eq:fric_dyn}, enclosed in $\tau_{\rho}$. 
To measure this parameter we fitted the intermediate scattering function following Ref.~\citenum{gallo1996}:
\begin{equation}
    F(q,t) = [1-A(q)]e^{-(t/\tau_s)^2} + A(q)e^{-(t/\tau_l)^{\gamma}},
    \label{eq:FqtFIT}
\end{equation}
considering two characteristic time-scales: at short times with $\tau_{\beta}=\tau_s \, \Gamma(1/2)/2$ and at long times with $\tau_{\alpha} = \tau_l \, \Gamma(1/\gamma) / \gamma $, where $\Gamma(x)$ is the Euler function. 
$\tau_{\rho}$ is then defined as the integral of Eq.~\eqref{eq:FqtFIT}, i.e. $\tau_{\rho}=(1-A(q))\tau_{\beta}+A(q)\tau_{\alpha}$.
We found that $\tau_\alpha$ and $\tau_\beta$ were similar at high temperature, but while $\tau_\beta$ remained constant with $T$, $\tau_\alpha$ exponentially increased when lowering $T$, becoming the main contribution to $\tau_{\rho}$ in the supercooled regime (see the ESI).
Overall, $\tau_{\rho}$ data are well described by a VTF law, analogous to friction, showing that the density relaxation is the main interfacial molecular mechanism that controls friction's temperature evolution. 
With that regard, in previous work on bulk supercooled liquids \cite{yamamoto1998,Kumar2007,Jeong2009,xu2009,Ikeda2011,shi2013,Kawasaki2017}, it is not obvious what time should the viscosity be related to; 
usually, only $\tau_\alpha$ is considered, and often an effective $\tau_\alpha$ is defined as the time for which the self or coherent intermediate scattering function equals $1/e$. 
For friction however, it is clear in the derivation of Eq.~\eqref{eq:fric_dyn} that the total relaxation time $\tau_\rho$ should be used \cite{govind2019}, and indeed, Eq.~\eqref{eq:fric_dyn} predicted correctly the relative temperature evolution of $\lambda$ only when using $\tau_\rho$ (see the ESI, where large differences between the different relaxation times are reported).  
Note that Eq.~\eqref{eq:fric_dyn} failed to reproduce $\lambda$ quantitatively; this is reminiscent of similar quantitative discrepancies reported in previous work using analogous approximations of the full Green-Kubo expression of $\lambda$ \cite{falk2010,Tocci2020}. 
We can compare the relaxation dynamics in our work with
experiments in bulk and confined supercooled water. 
For water in contact with graphene and LJ walls as well as for bulk water,
we predict a value of $\tau_\rho \sim 3$\,ps at 240\,K, see the ESI.
Experiments in bulk water~\cite{Cerveny2016,dehaoui2015}
report values between 20 and 30\,ps whereas
neutron scattering experiments in silica 
nanopores report relaxation times of about 100\,ps
at similar temperatures \cite{chen2006},
indicating that under confinement the relaxation dynamics 
can be slowed down dramatically at interfaces
where water dissociation and hydrogen bonding with the surface can occur,
as opposed to atomically smooth surfaces, such as those considered here.

\begin{figure}
    \centering
    \includegraphics[width=\linewidth]{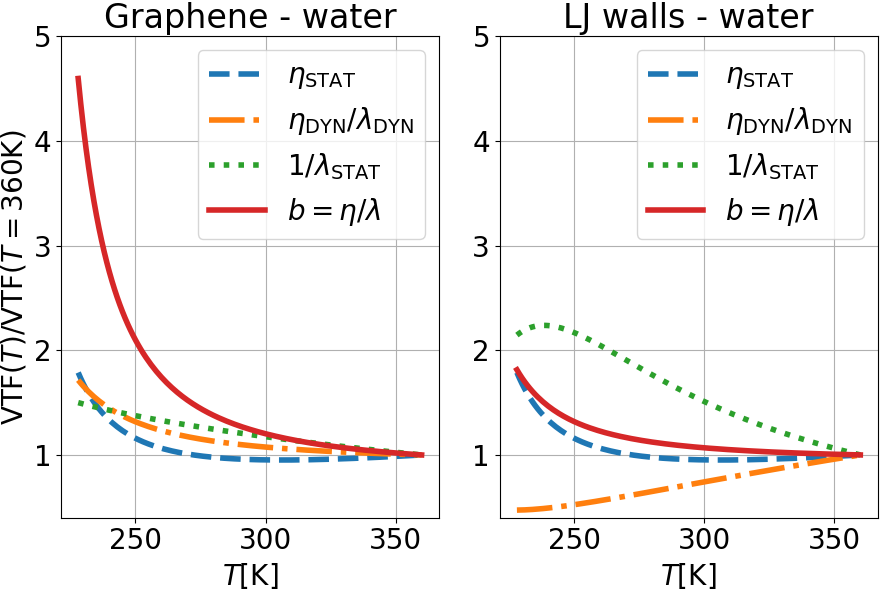}
    \caption{Temperature evolution of the static and dynamical contributions to the slip length $b=\eta/\lambda$ of water on graphene and LJ walls, normalized by the values at $360$\,K. The lines were obtained from VTF fits of the simulation results, see text for details.} 
    \label{fig:vtfratios}
\end{figure}

To then understand the temperature dependence of the slip length $b=\eta/\lambda$, we will decompose the viscosity into a static and a dynamical part in the same manner as for the friction coefficient: $\eta = \eta_\text{STAT} \times \eta_\text{DYN}$, with $\eta_\text{DYN} = \tau^{\text{bulk}}_{\rho} / (2k_\text{B}T)$ -- in analogy with the definition of $\lambda_\text{DYN}$, and with $\eta_\text{STAT} = \eta/\eta_\text{DYN}$.
The slip length can then be decomposed as follows: 
\begin{equation}
    b = \frac{\eta}{\lambda} =  \eta_{\mathrm{STAT}} \cdot \frac{\eta_{\mathrm{DYN}}}{\lambda_{\mathrm{DYN}}} \cdot \frac{1}{\lambda_{\mathrm{STAT}}}.
    \label{eq:b_decomposed}
\end{equation}
Figure~\ref{fig:vtfratios} illustrates the temperature evolution of the three contributions to $\lambda$ for water on LJ walls and graphene. In this figure, the lines are obtained from the ratios between VTF fits of the simulation results for $\eta$, $\lambda$, $\tau_\rho$ and $\tau_\rho^\text{B}$: specifically, $\eta_\text{STAT} \propto T\eta/\tau^{\mathrm{B}}_{\rho}$, $\eta_\text{DYN}/\lambda_\text{DYN} = \tau_\rho^\text{B} / \tau_\rho$, and $\lambda_\text{STAT} \propto T\lambda/\tau_{\rho}$. 
One can observe in Fig.~\ref{fig:vtfratios} that $1/\lambda_{\mathrm{STAT}}$ increases when decreasing $T$ for both interfaces. 
As detailed in the ESI, the stronger temperature variation of $\lambda_{\mathrm{STAT}}$ for the LJ walls can be related to the larger extension of the density profiles toward the wall at high temperatures. 
In bulk, $\eta_{\mathrm{STAT}}$ remains constant at high $T$, but it increases significantly when water enters its supercooled regime, for $T < 273$\,K, becoming the main contribution to $b$ significant increase in the same $T$ region. 
As a side note, following our choice to include $1/(k_\text{B}T)$ in $\eta_{\mathrm{DYN}}$, the fact that  $\eta_{\mathrm{STAT}}$ is constant in the liquid state corresponds to $\eta \propto \tau_\rho^\text{B}/T$; we suggest this correlation could replace more traditional ones used when studying supercooled liquids, $\eta \propto \tau_\alpha$ or $\eta \propto T \tau_\alpha$ \cite{yamamoto1998,Kumar2007,Jeong2009,xu2009,shi2013,Kawasaki2017}. 

Finally, to understand the relative increase of $b$ by $\sim 2$ times for the LJ wall and by $\sim 5$ times for graphene, we looked at the dynamic ratio $\eta_{\mathrm{DYN}}/\lambda_{\mathrm{DYN}}$. In Fig.~\ref{fig:vtfratios}(right), one can see that for LJ walls the interface relaxation time increases more when decreasing $T$ than the bulk one, compensating the static contribution and resulting in a smaller $b$ variation. 
In contrast, for graphene, due to the surface smoothness, there is no contribution from the wall to the slowing down of the interface dynamics with $T$ when compared to the bulk dynamics. 
Therefore, as for the temperature dependence of $\lambda$, we conclude that also with regard to $b$ it is not the different interfacial structures which contribute to its $T$ evolution but the different dynamics.

Before concluding, we would like to comment on a prediction for the temperature dependence of $b$ introduced by Bocquet and Barrat \cite{bocquet2007}, who wrote that $b$ should be proportional to $(k_\text{B}T)^2/\lambda_\text{STAT}$, in contrast with our results. This formula can be derived from Eq.~\eqref{eq:fric_dyn} by relating the density relaxation time $\tau_\rho$ to the collective diffusion coefficient $D_{q_{\parallel}}$: $\tau_\rho = 1/(q_{\parallel}^2 D_{q_{\parallel}})$, and by identifying $D_{q_{\parallel}}$ with the self-diffusion coefficient $D_0$, itself related to the viscosity through the Stokes-Einstein relation: $D_0 \propto k_B T / \eta$. However, while we found that indeed $D_{q_{\parallel}} \simeq D_0$ at room temperature, their temperature evolution is quite different, especially in the supercooled regime (see the ESI). 
Indeed, both diffusion coefficients arise from processes that happen at different scales and their relation is non-trivial: while $D_{q_{\parallel}}$ is related to collective diffusion in the sense that it comes from the density Fourier transform integration to all atoms positions, $D_0$ is referred to the diffusion of one molecule of $\sigma_l$ size.   

\paragraph*{Conclusions--} 

In this work we investigated the temperature evolution of bulk and interfacial hydrodynamic transport coefficients for water and MeOH confined between LJ walls and graphene. 
For a given liquid, the temperature evolution of viscosity and friction were described by the same laws, although with different parameters. 
The temperature evolution of interfacial friction was weakly affected by the wall type, but changed significantly with the liquid type. 
We then compared the temperature evolution of  viscosity $\eta$ and friction coefficient $\lambda$ by considering their ratio, defined as the slip length $b = \eta/\lambda$. We observed, from higher to lower $T$, that both transport coefficients evolved similarly 
in the high temperature region where the liquid is stable, but that for water, viscosity increased faster than friction in the supercooled regime, implying a fast growing slip length. The largest temperature effect on $b$ was observed for water and graphene walls, where it grew by a factor of $5$ from the highest to the lowest simulated temperature (225\,K), reaching a maximum value of $\sim 230$\,nm.

In order to understand the molecular mechanisms that control friction, we decomposed the friction coefficient $\lambda$ into the product of a static  contribution $\lambda_\text{STAT}$ and a dynamical one $\lambda_\text{DYN}$, in the form of an interface density relaxation time $\tau_\rho$. 
We observed a small variation of the static part with $T$, but the main contribution to the temperature dependence of friction came from the dynamical term.
Finally, in order to explain the temperature dependence of the slip length $b=\eta/\lambda$, we also decomposed the viscosity $\eta$ into a static term $\eta_\text{STAT}$ and a dynamical term $\eta_\text{DYN}$, controlled by the bulk density relaxation time $\tau_\rho^\text{B}$. The slip length could then be decomposed into three contributions: first, the interfacial static contribution $1/\lambda_\text{STAT}$; second, the bulk static contribution, $\eta_\text{STAT}$; and third, the relation between the bulk and interfacial dynamical terms $\eta_\text{DYN}/\lambda_\text{DYN} = \tau_\rho^\text{B} / \tau_\rho$. 
We observed that the viscosity static part, while it remained constant at high temperature, increased significantly in the supercooled regime, representing a major contribution to the slip length temperature evolution. 
We could finally relate the different slip length temperature dependence on LJ walls and graphene to the difference in interfacial dynamics on these two surfaces. %

We suggest that the promising predictions presented here should be within reach of experimental verification, with the recent accurate characterization of liquid-solid slip on new 2D materials and their derivative \cite{Secchi2016,Yang2017,Tunuguntla2017,Xie2018}, and investigation of supercooled water dynamics down to very small temperatures \cite{buchsteiner2006water,cerveny2010dynamics,chen2006,sjostrom2010dielectric,Jung2012,dehaoui2015,Cerveny2016}, e.g. $\sim 230$\,K in bulk \cite{dehaoui2015} and $\sim 130$\,K in confinement \cite{sjostrom2010dielectric}.
Moreover, beyond liquid-solid slip, many other new behaviors could arise in the promising field of supercooled nanofluidics. 
Potential venues to be explored
may include the study of contact angle
hysteresis and the flow in Wenzel and Cassie-Baxter states in nanopatterned surfaces under supercooling. Analogous phenomena have already been explored experimentally in the context of anti-icing surfaces \cite{Kreder2016}. Thus, it
stands to reason that the study of interfacial water
transport in the supercooled regime
may also be relevant to advance  understanding of heterogeneous ice nucleation
and to improve strategies to design
anti-freezing coatings \cite{Sosso2016}. 
Overall we hope the findings obtained here by investigating water friction as a function of temperature down to the supercooled regime 
will help understanding generally the molecular mechanisms underlying both interfacial and bulk hydrodynamic transport in this fascinating liquid and motivate experimenters to find protocols to measure
water slippage under supercooling.

\begin{acknowledgments}
The authors thank Li Fu, Simon Gelin, Yasutaka Yamaguchi, Takeshi Omori, Emmanuel Guillaud, Bruno Issenmann for fruitful discussions. 
We are also grateful for HPC resources
from GENCI/TGCC (grants A0050810637 and A0070810637), and 
from the PSMN mesocenter in Lyon. 
This work is supported by the ANR, Project ANR-16-CE06-0004-01 NECtAR. LJ is supported by the Institut Universitaire de France. GT   is   supported   by   the   Swiss National Science Foundation through the  project PZ00P2\_179964.
\end{acknowledgments}

%

\end{document}